\documentclass{Interspeech2024_arxiv}
\usepackage{balance}
\usepackage{url}
\usepackage{siunitx}
\usepackage{multirow}
\usepackage{amsmath}
\usepackage{float}
\usepackage{multirow}

\usepackage[utf8]{inputenc}
\usepackage[T1]{fontenc}

\newcommand{\FO}{$F_0$\xspace}

\usepackage{pdfrender}
\newcommand\tablebf[1]{\textpdfrender{TextRenderingMode=FillStroke,LineWidth=0.5}{#1}}

\interspeechcameraready

\title{Voice Conversion-based Privacy through Adversarial Information Hiding}

\name[affiliation={1}]{Jacob J}{Webber}
\name[affiliation={2}]{Oliver}{Watts}
\name[affiliation={3}]{Gustav Eje}{Henter}
\name[affiliation={4}]{Jennifer}{Williams}
\name[affiliation={1}]{Simon}{King}

\address{
  $^1$The University of Edinburgh, UK;
  $^2$SpeakUnique Ltd., UK; \\
  $^3$KTH Royal Institute of Technology, Sweden;
  $^4$University of Southampton, UK
\email{\href{mailto:j.j.webber@ed.ac.uk}{j.j.webber@ed.ac.uk}, \href{mailto:simon.king@ed.ac.uk}{simon.king@ed.ac.uk}}}


\begin{document}

\maketitle

\begin{abstract}

Privacy-preserving voice conversion aims to remove only the attributes of speech audio that convey identity information, keeping other speech characteristics intact. This paper presents a mechanism for privacy-preserving voice conversion that allows controlling the leakage of identity-bearing information using adversarial information hiding. This enables a deliberate trade-off between maintaining source-speech characteristics and modification of speaker identity. As such, the approach improves on voice-conversion techniques like CycleGAN and StarGAN, which were not designed for privacy, meaning that converted speech may leak personal information in unpredictable ways. Our approach is also more flexible than ASR-TTS voice conversion pipelines, which by design discard all prosodic information linked to textual content. Evaluations show that the proposed system successfully modifies perceived speaker identity whilst well maintaining source lexical content.


\end{abstract}

\section{Introduction}
As the ability of state-of-the-art systems to extract sensitive information from speech signals has increased, so has the importance of speech privacy.
Speech privacy has become a prominent research area since the VoicePrivacy Initiative was introduced in 2018 \cite{voiceprivintroducing}. This initiative was followed by the establishment of biennial VoicePrivacy Challenges\footnote{\url{https://www.voiceprivacychallenge.org/}}\cite{voiceprivresults}. A speaker's identity is deeply entangled among all other acoustic and semantic characteristics of speech~\cite{williams2019disentangling}. It is therefore challenging to remove speaker-identifying information without destroying other components of the speech signal. While the VoicePrivacy Challenge systems and evaluation campaigns have been exploring the trade-offs between retaining source speech attributes while simultaneously preserving privacy, we introduce a nuanced and principled attempt to achieve \emph{differential privacy} -- which we define as the ability to select or control how much information is hidden or concealed. As more national and international bodies consider regulating both privacy and the use of artificial intelligence~\cite{mourby2018pseudonymised,voigt2017eu}, privacy solutions must keep pace with these changing needs 

Considering how rich the speech signal is with information, there are may different ways to view voice privacy beyond traditional definitions. These include neutralising prosody or other speaking patterns that could identify a person, removing aspects of speech that convey a health condition, transforming the age or gender of the speaker~\cite{adversarial_dis,benaroya2021voice}, and concealing or masking targeted words~\cite{10363030}. Voice privacy can also be viewed from within a paradigm of how speech data is transmitted between humans or machines, and how that information may be intercepted~\cite{williams22_spsc}. 
Controllability is key to developing solutions that can address multiple types of privacy. We introduce a new mechanism for controlling leakage of identity-bearing information using adversarial information hiding that deliberately balances the trade-offs between maintaining source-speech characteristics and modifications of speaker identity. We draw from lessons learned in voice conversion techniques like CycleGAN~\cite{kaneko2019cyclegan} and StarGAN~\cite{kaneko2019stargan} which were not designed specifically with privacy in mind. This means that converted speech may leak personal information in unpredictable ways. 

Our contributions in this paper address the idea of controllable privacy in a new way. We build on work introduced with the Hider-Finder-Combiner (HFC-\FO)~\cite{hfc} model, which in turn has some similarities to Fader Networks~\cite{fadernetworks}. The HFC-\FO model was devised to modify \FO contours of source speech, a control parameter which, like speaker identity, is deeply entangled with all other aspects of a speech signal~\cite{williams2019disentangling}.
%
We introduce HFC-VP as a system that creates a disentangled hidden representation which can then be recombined with an arbitrary speaker embedding. To the best of our knowledge, the techniques that we introduce represent the first effort to achieve anonymization solely through minimising the mutual information between a predicted probability distribution of speaker identity and a prior distribution of speaker identity, from a large dataset of 904 speakers. Our principled and ``unguided'' approach means that our single system is capable of modifying different aspects of speech, such as prosody or lexical content, exactly where it is necessary to preserve privacy.

Our work offers substantial changes to the original HFC-\FO system that was presented in \cite{hfc}. We have adapted the training losses (Section \ref{sec:losses}) in a manner that allows us to re-purpose the underlying architecture for the task of controllable speaker privacy. Our system also replaces the component networks, which now use powerful transformer encoders and residual convolutional networks in place of RNNs. The resulting system outperforms a strong neural source-filter baseline in our evaluations. It also provides the possibility of sharing an anonymised speech representation that can be resynthesised at a later date.
Readers are encouraged to listen to our audio examples\url{https://hfcvp.github.io/}.

\section{Background}




There is increasing interest in using speech signal disentanglement and privacy control with speaker anonymisation~\cite{disentangledallyouneed}. Systems have applied inductive biasing, using models that only have the power to learn specific attributes. For example, bottlenecks have been used to guide synthesis by accounting for features that are desirable to preserve~\cite{qian2020unsupervised, amazon_disentangle,disentangle_luck,adversarial_dis,nvidia_voice_conversh}. An example is preserving the lexical content of utterances alongside such bottleneck features. Other privacy-preserving approaches~\cite{adversarial_dis} have applied Fader Networks~\cite{fadernetworks} to modify specific scalar attributes of the speech signal.


The Hider-Finder-Combiner (HFC) architecture was first introduced in \cite{hfc} for the purpose of performing general modifications to the speech signal. The original paper showed that the system of training worked well for \FO modification. We significantly extend their architecture (Section~\ref{sec:method}) and also show how our extensions can adapt the architecture more specifically for preserving speaker privacy.

The HFC system equates the problem of speech signal modification with \emph{disentanglement}~\cite{williams2019disentangling,qian2020unsupervised}. Parameterising certain aspects of speech signals (which are called \emph{control parameters}) is challenging because the input to machine learning-based speech modification systems (e.g., mel spectrograms) contains information relating to other aspects of the speech signal that may be desirable to change as well as aspects that may be desirable to preserve. The HFC system overcame this problem by using adversarial information hiding.


HFC employs a \emph{Hider} network to generate a latent representation, $h$, which contains as little information as possible about the control parameter, but retains all \textit{other} information from its input so that, when combined with the control parameter, in this case another network, the \emph{Combiner} can reconstruct the original signal. HFC also uses a \emph{Finder} adversary to ensure that the Hider network does not leak information relating to the control parameter. This Finder predicts the conditional probability distribution given $h$:
\begin{align}
h & = H(x) \\
x & \approx C(h,c) \\
F(h) & \approx p(c|h)
\end{align}
where $c$ is the class assigned to the control parameter. In the original \cite{hfc} the control parameter is a time sequence of binned \FO values. The HFC architecture is shown in figure \ref{fig:adv}.

The way that HFC is trained is similar to a typical \emph{Generative Adversarial Network} (GAN), in which a generator network aims to fool a discriminator (adversary) network into mis-classifying its output (e.g., to treat output as genuine \cite{gan}). In HFC, as in GANs, the generators (Hider and Combiner together) and the discriminator (Finder) are trained in turn. We provide a more detailed description of the original training losses, as well as how we have modified these losses for our system in Section \ref{sec:hfc_2}. Fader networks are somewhat similar to HFC, although simpler in that they only attempt to disentangle a single scalar value.

We also note \cite{chou18_interspeech}, which uses a similar three-component network to HFC and Fader Networks, applying these to the task of voice conversion. However, this work differs from that presented here in that it takes a simpler approach compared with our leakage losses discussed in section \ref{sec:losses}, uses RNN-based component networks rather than a transformer as described in section \ref{sec:component_networks}, is not evaluated on a privacy-related task and does not generalise to unseen target speakers.

\section{Method}
\label{sec:method}

\subsection{Losses} 
\label{sec:losses}

In \cite{hfc} three losses had been proposed. A Finder loss trains the adversary, while the hider and combiner, trained alternately with the finder, use a combiner loss and a leakage loss, which measure reconstruction quality and information leakage respectively.
This leakage loss is defined as $\text{Var}[F(c|H(x))]$, i.e. the variance of the conditional probability predicted by the finder.

While the proposed losses from \cite{hfc} were useful for their goal of F0 modifications, they are not well-suited for our task of differential privacy. Therefore, in this paper we contribute a fundamental change to the loss functions in two ways to achieve more consistency. We adopt an information theoretic approach, describing the leakage loss as a measure of information leakage between a predicted distribution, and that of a \emph{prior} distribution, where this prior is the general distribution of the underlying data. The original HFC used a mean squared error (MSE) to measure the leakage but an entropy-based approach to train the finder. 

By changing the loss of both leakage and finder to MSE, we then measure the divergence between two probability distributions with a theoretical basis~\cite{pearson}, with the understanding that this approach can be useful for GAN training~\cite{lsgan}.

\begin{equation}
    \mathcal{L}_{\text{leakage}} = \frac{1}{C} \sum_c (F(c|H(x)) - p(c) ) ^2
    \label{eq:mse_l}
\end{equation}

\begin{equation}
   \mathcal{L}_{\text{finder}} = \frac{1}{C} \sum_c (F(c|H(x)) - I(c|x)) ^2
   \label{eq:mse_f}
\end{equation}

It is alternatively possible to use a more information-theoretic approach. Ignoring constant terms

\begin{align}
    \mathcal{L}_{\text{finder}} &= D_{KL}(p(c|x) || F(c|H(x))) \\
                      &= - \log(F(c_i|H(x_i))
\end{align}

\begin{align}
    \mathcal{L}_{\text{leakage}} &= D_{KL}(p(c) || F(c|H(x))) \\
                       &= - \sum_c p(c) \log F(c| H(x))
\end{align} 
where $c$ is the class (speaker identity in our case) and $x$ is the input mel spectrogram. We optimise $H$ to minimise $\mathcal{L}_{\text{leakage}}$ and $F$ to minimise $\mathcal{L}_{\text{finder}}$. Both $H$ and $C$ are optimised to minimise a MSE reconstruction loss, $\mathcal{L}_{\text{combiner}}$. We weight these terms by deriving a total loss with which to optimise both $H$ and $C$
$$\mathcal{L}_{G} = \mathcal{L}_{\text{combiner}} + \beta \mathcal{L}_{\text{leakage}}$$

For this work we exclusively use our new losses defined in \ref{eq:mse_l} and \ref{eq:mse_f}. We provide implementations of the other loss regime in our code release. We note that $\mathcal{L}_{\text{leakage}}$ defined in equation \ref{eq:mse_l} is equivalent to the original HFC variance loss when the prior distribution is uniform. This prior can be calculated by finding the softmax of the histogram of the dataset across classes.
 
\begin{table*}[ht!]
\caption{Objective results comparing system performance across tasks for retained lexical content (WER), emotion recognition (UAR), and speaker anonymisation (EER). Down arrows ($\downarrow$) and up arrows ($\uparrow$) indicate whether lower or higher values represent better or worse performance, respectively. For anonymisation, higher EER is preferred.}
\begin{center}
\begin{tabular}{@{}l| c c | c c | c c c c@{}} 
 \toprule
 \multirow{2}{*}{Task} & \multicolumn{2}{c|}{No Anonymisation} & \multicolumn{2}{c|}{VPC Baselines} & \multicolumn{4}{c@{}}{HFC-VP}  \\ 
  & Original & HiFiGAN & B1 & B2 & $\beta=.050$ & $\beta=.060$ & $\beta=.065$ & $\beta=.070$  \\ 
 \midrule
 WER Libri Test $\downarrow$&  1.839 & 2.024 & 6.253 & 10.381 & 30.976 & 3.572 & 6.024 & \tablebf{3.343} \\  
 \midrule
 UAR IEMOCAP Test $\uparrow$ & 71.062 & 56.656 & 42.314 & \tablebf{53.491} & 46.785 & 49.416 & 47.104 & 48.444  \\
 \midrule
 EER OA m Libri Test $\uparrow$ & 0.418 & 2.229 & N/A & \tablebf{28.508} & 18.485 & 14.053 & 17.817 & 12.697 \\
 EER OA f Libri Test $\uparrow$ & 8.761 & 8.941 & N/A & \tablebf{33.210} & 19.195 & 12.998 & 18.428 & 11.496 \\
 EER AA m Libri Test $\uparrow$ & 0.418 & 0.667 & \hphantom{0}6.682 & \tablebf{26.988} & 17.775 & \hphantom{0}9.134 & 15.366 & \hphantom{0}7.127 \\
 EER AA f Libri Test $\uparrow$ & 8.761 & 8.761 & 10.584 & \tablebf{32.118} & 16.104 & 11.694 & 14.597 & 10.218 \\
 \bottomrule
\end{tabular}
\end{center}
\label{tab:obj}
\vspace{-\baselineskip}
\end{table*}

\subsection{Data}
We use the train-clean-360 split of the LibriTTS dataset \cite{libritts} for training. The dataset contains 360 hours of read speech audio at \SI{22.05}{\kilo\hertz}. The dataset contains 904 speakers. HFC-VP synthesises mel spectrograms which are converted into the audio domain using a pretrained SpeechBrain\cite{speechbrain} implementation of HiFiGAN. 
At training time the system uses both speaker ID and speaker embeddings extracted with SpeechBrain's ECAPA-TDNN, pretrained on the Voxceleb dataset. Because ECAPA-TDNN embeddings are calculated from mel spectrograms, it is possible to calculate these embeddings directly from HFC-VP's output without vocoding, which could be useful for future experiments.

\subsection{Model}
\label{sec:hfc_2}

\begin{figure}[ht]
    \centering
    \includegraphics[width=\linewidth]{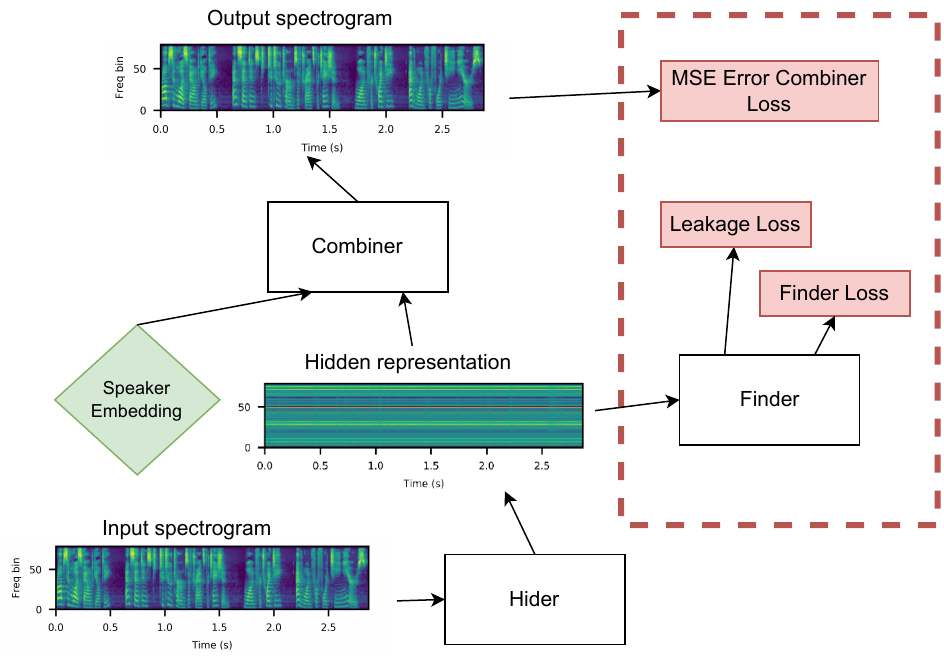}
    \caption{Adversarial training setup.}
    \label{fig:adv}
\end{figure}

The training of the system is largely unmodified from \cite{hfc}, except the control parameter is changed from \FO to speaker identity. We use two representations of speaker identity. For the finder loss we use the index of the speaker in the training set. For the combiner, the input control parameter takes the form of a pretrained speaker embedding, specifically the ECAPA mel-spectrogram embeddings from SpeechBrain\footnote{\url{https://huggingface.co/speechbrain/spkrec-ecapa-voxceleb-mel-spec}}. The use of mel spectrogram-derived embeddings means it should be possible to calculate embeddings directly on the output of HFC, without using a vocoder. Such experiments are left to future work. The training setup is shown in figure \ref{fig:adv}. As the finder is not used during inference, HFC can generalise to unseen speakers.

\subsection{Component networks}
\label{sec:component_networks}

\textbf{Hider:}
Unlike the hider network from HFC-\FO, which uses an RNN, the proposed hider network is based on the generator residual blocks from HiFiGAN V2 \cite{hifigan} without the upsampling blocks. The 80 input mel spectrogram features per time step are treated as input channels to a 1D convolution of width 7 which increases the number of channels to 128. This is passed into 3 residual blocks, each containing 3 1D convolutions of kernel sizes 3, 7, 11 and dilations
1, 3 and 5 respectively. The residual blocks use a leaky-ReLU discontinuity and weight-norm. The output of these residual blocks is reduced to 80 channels (the dimensionality of our hidden representation) using another 1D convolution with kernel width 7. Our hidden representation is a time-sequence with the same duration as the input mel spectrogram. A true seq-to-seq approach with a fixed size hidden representation is an interesting avenue to explore in future work as it allow for concealment of speaker-specific duration patterns.
%

\textbf{Finder:}
The finder network is based on that used in \cite{hfc}. The hidden representation is passed into a 1D convolution with kernel size 9. The convolution is applied along the width dimension rather than the time dimension. This is passed into a 3 layer GRU of size 200. A final linear layer maps the output to the number of speakers in the dataset. Hyper-parameters (model sizes and finder learning-rate) were found by sweeping, training the finder on ground truth mel-spectrograms as a speaker classification model. On the LibriTTS train-clean-100 dataset the model achieved \SI{97.3}{\percent} speaker classification accuracy. However, the finder was trained from scratch for the HFC training.

\textbf{Combiner:}
The combiner network is completely changed from HFC-\FO. We use a model based on FastPitch 2. We remove the token encoder transformer. Instead a linear layer maps the 80 hidden features to 384 features. The speaker embedding is projected from 192 features to 384 using a linear layer and then added to the hidden projection. 
%
We remove the $F_0$ and energy projections; incorporating these is left as future work.
A four-layer transformer encoder network is used followed by a convolutional postnet. During training, the output of both the transformer and the postnet are trained using the combiner loss. The hyper-parameters for the postnet and transformer are as in~\cite{fastspeech2}, specifically the SpeechBrain implementation.

\subsection{Hyper-parameters}
Initial learning-rates were selected using PyTorch Lightning's automatic learning-rate selection. In the case of the hider this was used with training the finder as a speaker identity classifier on ground truth mel spectrograms. For the $\text{loss}_G$ this was found by performing a copy synthesis task 
using just the combiner on mel spectrograms. Exponential learning-rate decay was used after the first 100 epochs with $\gamma=0.999$. An initial range for $\beta$ was chosen using informal experimentation. For $\beta < \num{5e-2}$ the control did not significantly affect speaker identity. For $\beta > \num{7e-2}$ training did not converge on acceptably high quality audio output. The hider, finder and combiner networks contain $2.2$, $0.87$ and $23$ million trainable parameters respectively.
Samples, source code and trained model files are available online\footnote{\url{https://github.com/jacobjwebber/hider_finder_combiner}}

\subsection{Anonymisation}
Our system allows the specification of a target speaker through a pre-trained speaker embedding. The choice of which embedding to use is non-trivial. Some users requiring anonymisation may want to speak with a consistent voice between utterances, whereas others may not want to be easily associated with speech that they have previously shared. For our experiments we take the \emph{utterance} level approach, such that each utterance is anonymised to a random external speaker using a pool of speaker embeddings from VCTK \cite{vctk}. We choose this dataset due to its high quality and to verify that HFC performs well with target speakers that are unseen at training time. We note that other approaches aim to maximise the difference between target and source speech embeddings, which could yield improved results in objective analysis, as cases where the target speaker and source speaker happen to be similar could be incorrectly identified as being poorly anonymised. We do not use that approach as it leaks information about the speaker, something we want to avoid. Speech anonymised in this way reveals characteristics that are likely to be \emph{opposite} to the source speaker, which could yield sensitive information through triangulation. To evaluate the system in terms of \emph{utility} and \emph{privacy} we use the baseline and benchmarks provided as part of the VoicePrivacy Challenge (VPC) 2024, which are freely available\footnote{\url{https://github.com/Voice-Privacy-Challenge/Voice-Privacy-Challenge-2024}}.

\section{Results}

\begin{table}[]
    \caption{Losses for Combiner and Leakage modules change based on different values of $\beta$. Both decrease as $\beta$ increases.}    \centering
    \begin{tabular}{@{}l|c c c c@{}}
         \toprule
         Loss & $\beta=.050$ & $\beta=.060$ & $\beta=.065$ & $\beta=.070$  \\
         \midrule
         Combiner &  0.02962\hphantom{0} & 0.02499\hphantom{0} & 0.02677\hphantom{0} & 0.02475\hphantom{0}\\
         Leakage & 0.005686 & 0.004225 & 0.001497 &  0.001379\\
         \bottomrule
    \end{tabular}
    \label{tab:beta}
    \vspace{-\baselineskip}
\end{table}

A key hyper-parameter is the choice of the value $\beta$ used for balancing the importance of the adversarial information-hiding loss with the reconstruction loss. Results of the losses with varying values of this hyper-parameter are shown in Table~\ref{tab:beta}. Interestingly these show that \emph{both} losses reduce with $\beta$. This differs from an expectation that $\beta$ parameterises a trade-off between these two variables. It was not possible to achieve stable training with $\beta > 0.07$.


The objective evaluation approaches are taken from the VPC 2024 evaluation plan \cite{tomashenko2024voiceprivacy}, using their freely available implementation. We evaluate 4 HFC systems with varying values of beta and baselines B1 and B2 from the VPC 2024 evaluation plan. We also include HiFiGAN copy synthesis from ground truth mel spectrograms, with no attempt at anonymisation. Baseline B1 uses a neural source-filter approach. Baseline B2 uses signal processing to hide speaker identity, specifically the McAdams coefficient, as described in \cite{mcadams}. Due to compute and time constraints, we were unable to recreate results with baseline B1, and therefore unable to verify or fill in missing results from Table~\ref{tab:obj}. 
We use Unweighted Average Recall (UAW) to evaluate the Speech Emotion Recognition (SER) performance of our systems, ensuring the emotional content of the source is preserved. An ASR-derived word error rate (WER) measures the retained lexical content. A speaker verification model measures an Equal Error Rate (EER), which is a measure of how well anonymised our output is. As with \cite{tomashenko2024voiceprivacy}, the systems are tested with both anonymised (AA) and original (OA) enrollment data. We do not finetune EER models on anonymised speech. The datasets used for evaluation are the IEMOCAP emotion recognition dataset and the LibriSpeech test set. We show these results in Table~\ref{tab:obj}. We are the first to show a principled trade-off between anonymization and speech utility measured.

The WER of all HFC systems, with the exception of $\beta = 0.05$ is better than B2, to the extent that a formal listening test for speech utility preservation is not necessary. In addition to the very strong WER score, we demonstrate the output quality of our system by making samples available online\footnote{\url{https://hfcvp.github.io/}}. 
This is in contrast to baseline system B2 which -- although it obtains very good scores for privacy -- imposes a high cognitive load on listeners due to its poor signal quality. Our proposed system outperforms the benchmark neural system B1 both in terms of privacy and utility.

A detailed analysis of computational performance is left as future work. However, as the largest component of HFC is a much reduced version of FastSpeech 2~\cite{fastspeech2}, our system is inevitably faster. The overwhelming majority of compute time used during inference is taken by the HiFiGAN vocoder.

\section{Conclusion}
\label{sec:conc}

We present a system that anonymises speech in a principled way, demonstrating strong performance that captures the trade-offs between speech utility and anonymisation performance with objective measures. 
As our finder is trained to classify between the 904 speakers of LibriTTS, the information that HFC-VP obscures is likely only that which is useful for categorizing these speakers. Other attributes (e.g., accent or prosody) may be maintained, particularly as the dataset does not seem to be particularly diverse in terms of speaker demographics. The study of which attributes are affected and which remain untouched opens interesting avenues of future research. It is also possible to modify features other than or in addition to speaker identity by adding classification heads to the finder, or by adding additional finders, as well as by adding extra feature embeddings to the internal representation of the FastSpeech2-style combiner. The ability to redact specific characteristics, with the only requirement being that the characteristic is labelled on some dataset, is an exciting feature of HFC-VP-type systems.



\balance
\bibliographystyle{IEEEtran}
\bibliography{mybib}

\begin{thebibliography}{10}
\providecommand{\url}[1]{#1}
\csname url@samestyle\endcsname
\providecommand{\newblock}{\relax}
\providecommand{\bibinfo}[2]{#2}
\providecommand{\BIBentrySTDinterwordspacing}{\spaceskip=0pt\relax}
\providecommand{\BIBentryALTinterwordstretchfactor}{4}
\providecommand{\BIBentryALTinterwordspacing}{\spaceskip=\fontdimen2\font plus
\BIBentryALTinterwordstretchfactor\fontdimen3\font minus \fontdimen4\font\relax}
\providecommand{\BIBforeignlanguage}[2]{{%
\expandafter\ifx\csname l@#1\endcsname\relax
\typeout{** WARNING: IEEEtran.bst: No hyphenation pattern has been}%
\typeout{** loaded for the language `#1'. Using the pattern for}%
\typeout{** the default language instead.}%
\else
\language=\csname l@#1\endcsname
\fi
#2}}
\providecommand{\BIBdecl}{\relax}
\BIBdecl

\bibitem{voiceprivintroducing}
N.~Tomashenko, B.~M.~L. Srivastava, X.~Wang, E.~Vincent, A.~Nautsch, J.~Yamagishi \emph{et~al.}, ``{Introducing the VoicePrivacy Initiative},'' in \emph{Proc. Interspeech 2020}, 2020, pp. 1693--1697.

\bibitem{voiceprivresults}
N.~Tomashenko, X.~Wang, E.~Vincent, J.~Patino, B.~M.~L. Srivastava, P.-G. No{\'e} \emph{et~al.}, ``{The VoicePrivacy 2020 Challenge: Results and findings},'' \emph{Computer Speech \& Language}, vol.~74, p. 101362, 2022.

\bibitem{williams2019disentangling}
J.~Williams and S.~King, ``Disentangling style factors from speaker representations.'' in \emph{Proc. of Interspeech 2019}, 2019, pp. 3945--3949.

\bibitem{mourby2018pseudonymised}
M.~Mourby, E.~Mackey, M.~Elliot, H.~Gowans, S.~E. Wallace, J.~Bell, H.~Smith, S.~Aidinlis, and J.~Kaye, ``{Are ‘pseudonymised’ data always personal data? Implications of the GDPR for administrative data research in the UK},'' \emph{Computer Law \& Security Review}, vol.~34, no.~2, pp. 222--233, 2018.

\bibitem{voigt2017eu}
P.~Voigt and A.~Von~dem Bussche, ``{The EU General Data Protection Regulation (GDPR)},'' \emph{A Practical Guide, 1st Ed., Cham: Springer International Publishing}, vol.~10, no. 3152676, pp. 10--5555, 2017.

\bibitem{adversarial_dis}
P.-G. Noé, A.~Emînî, M.~Driss, T.~Parcollet, A.~Nautsch, and J.-F. Bonastre, ``{Adversarial Disentanglement of Speaker Representation for Attribute-Driven Privacy Preservation},'' in \emph{Proc. Interspeech 2021}, 08 2021.

\bibitem{benaroya2021voice}
L.~Benaroya, N.~Obin, and A.~Roebel, ``Manipulating voice attributes by adversarial learning of structured disentangled representations,'' \emph{Entropy}, vol.~25, no.~2, p. 375, 2023.

\bibitem{10363030}
J.~Williams, K.~Pizzi, P.-G. Noe, and S.~Das, ``{Exploratory Evaluation of Speech Content Masking},'' in \emph{Speech Communication; 15th ITG Conference}, 2023, pp. 215--219.

\bibitem{williams22_spsc}
J.~Williams, K.~Pizzi, S.~Das, and P.-G. Noé, ``{New challenges for content privacy in speech and audio},'' in \emph{Proc. 2nd Symposium on Security and Privacy in Speech Communication}, 2022, pp. 1--6.

\bibitem{kaneko2019cyclegan}
T.~Kaneko, H.~Kameoka, K.~Tanaka, and N.~Hojo, ``{CycleGAN-VC2: Improved cyclegan-based non-parallel voice conversion},'' in \emph{IEEE International Conference on Acoustics, Speech and Signal Processing (ICASSP)}, 2019, pp. 6820--6824.

\bibitem{kaneko2019stargan}
------, ``{StarGAN-VC2: Rethinking Conditional Methods for StarGAN-Based Voice Conversion},'' \emph{Proc. of Interspeech 2019}, 2019.

\bibitem{hfc}
J.~J. Webber, O.~Perrotin, and S.~King, ``{Hider-finder-combiner: An adversarial architecture for general speech signal modification},'' in \emph{Proc. of Interspeech 2020}, 2020, pp. 3206--3210.

\bibitem{fadernetworks}
G.~Lample, N.~Zeghidour, N.~Usunier, A.~Bordes, L.~Denoyer \emph{et~al.}, ``Fader networks: Manipulating images by sliding attributes,'' in \emph{Advances in Neural Information Processing Systems}, 2017, pp. 5963--5972.

\bibitem{disentangledallyouneed}
P.~Champion, D.~Jouvet, and A.~Larcher, ``Are disentangled representations all you need to build speaker anonymization systems?'' in \emph{Proc. of Interspeech 2022}, 2022.

\bibitem{qian2020unsupervised}
K.~Qian, Y.~Zhang, S.~Chang, M.~Hasegawa-Johnson, and D.~Cox, ``Unsupervised speech decomposition via triple information bottleneck,'' in \emph{International Conference on Machine Learning}.\hskip 1em plus 0.5em minus 0.4em\relax PMLR, 2020, pp. 7836--7846.

\bibitem{amazon_disentangle}
J.~Mosiński, P.~Biliński, T.~Merritt, A.~Ezzerg, and D.~Korzekwa, ``{AE-Flow: Autoencoder Normalizing Flow},'' in \emph{IEEE International Conference on Acoustics, Speech and Signal Processing (ICASSP)}, 2023, pp. 1--5.

\bibitem{disentangle_luck}
C.~Peyser, W.~R. Huang, A.~Rosenberg, T.~Sainath, M.~Picheny, and K.~Cho, ``{Towards Disentangled Speech Representations},'' in \emph{Proc. Interspeech 2022}, 2022, pp. 3603--3607.

\bibitem{nvidia_voice_conversh}
S.~Kovela, R.~Valle, A.~Dantrey, and B.~Catanzaro, ``{Any-to-Any Voice Conversion with F0 and Timbre Disentanglement and Novel Timbre Conditioning},'' in \emph{IEEE International Conference on Acoustics, Speech and Signal Processing (ICASSP)}, 2023, pp. 1--5.

\bibitem{gan}
I.~Goodfellow, J.~Pouget-Abadie, M.~Mirza, B.~Xu, D.~Warde-Farley, S.~Ozair \emph{et~al.}, ``{Generative Adversarial Nets},'' in \emph{Advances in Neural Information Processing Systems 27}, 2014, pp. 2672--2680.

\bibitem{chou18_interspeech}
J.~chieh Chou, C.~chieh Yeh, H.~yi~Lee, and L.~shan Lee, ``{Multi-target Voice Conversion without Parallel Data by Adversarially Learning Disentangled Audio Representations},'' in \emph{Proc. Interspeech 2018}, 2018, pp. 501--505.

\bibitem{pearson}
K.~Pearson, ``X. on the criterion that a given system of deviations from the probable in the case of a correlated system of variables is such that it can be reasonably supposed to have arisen from random sampling,'' \emph{The London, Edinburgh, and Dublin Philosophical Magazine and Journal of Science}, vol.~50, no. 302, pp. 157--175, 1900.

\bibitem{lsgan}
X.~Mao, Q.~Li, H.~Xie, R.~Y. Lau, Z.~Wang, and S.~Paul~Smolley, ``Least squares generative adversarial networks,'' in \emph{Proceedings of the IEEE International Conference on Computer Vision}, 2017, pp. 2794--2802.

\bibitem{libritts}
H.~Zen, V.~Dang, R.~Clark, Y.~Zhang, R.~J. Weiss, Y.~Jia, Z.~Chen, and Y.~Wu, ``{LibriTTS: A Corpus Derived from LibriSpeech for Text-to-Speech},'' in \emph{Proc. Interspeech 2019}, 2019, pp. 1526--1530.

\bibitem{speechbrain}
M.~Ravanelli, T.~Parcollet, P.~Plantinga, A.~Rouhe, S.~Cornell, L.~Lugosch \emph{et~al.}, ``{SpeechBrain}: A general-purpose speech toolkit,'' 2021, arXiv:2106.04624.

\bibitem{hifigan}
J.~Kong, J.~Kim, and J.~Bae, ``{HiFi-GAN}: Generative adversarial networks for efficient and high fidelity speech synthesis,'' in \emph{Proc. NeurIPS}, vol.~33, 2020, pp. 17\,022--17\,033.

\bibitem{fastspeech2}
Y.~Ren, C.~Hu, X.~Tan, T.~Qin, S.~Zhao, Z.~Zhao, and T.-Y. Liu, ``{FastSpeech 2: Fast and High-Quality End-to-End Text to Speech},'' in \emph{International Conference on Learning Representations}, 2021.

\bibitem{vctk}
C.~Veaux, J.~Yamagishi, and S.~King, ``{The voice bank corpus: Design, collection and data analysis of a large regional accent speech database},'' in \emph{2013 International Conference Oriental COCOSDA held jointly with 2013 Conference on Asian Spoken Language Research and Evaluation (O-COCOSDA/CASLRE)}, 2013, pp. 1--4.

\bibitem{tomashenko2024voiceprivacy}
N.~Tomashenko, X.~Miao, P.~Champion, S.~Meyer, X.~Wang, E.~Vincent, M.~Panariello, N.~Evans, J.~Yamagishi, and M.~Todisco, ``The voiceprivacy 2024 challenge evaluation plan,'' \emph{arXiv preprint arXiv:2404.02677}, 2024.

\bibitem{mcadams}
J.~Patino, N.~Tomashenko, M.~Todisco, A.~Nautsch, and N.~Evans, ``{Speaker Anonymisation Using the McAdams Coefficient},'' in \emph{Proc. of Interspeech 2021}, 2021, pp. 1099--1103.

\end{thebibliography}

\end{document}